%% file: main.tex
\documentclass{nature}

\usepackage[top=0.8in, bottom=1in, left=0.8in, right=0.8in]{geometry}
\usepackage[numbers]{natbib}

\usepackage{graphicx}
\usepackage{amsmath}
\usepackage{amssymb}
\usepackage{threeparttable} 
\usepackage{caption}
\usepackage{bibunits}
\usepackage[dvipsnames]{xcolor} 
\usepackage{adjustbox}
\usepackage{xurl}
\usepackage{soul}
\usepackage{comment}
\usepackage[normalem]{ulem}

\defaultbibliography{sn-bibliography.bib}
\defaultbibliographystyle{naturemag}

\title{Efficient Interstellar Grain Growth from High Sticking Coefficients on Amorphous Carbon Dust}

\author{
Clarke J. Esmerian$^{1}$\textsuperscript{*}, Duncan Bossion$^{2}$\textsuperscript{*}, Francois Dulieu$^{3}$, Saoud Baouche$^{3}$, Alexey Potapov$^{4}$, W. M. C. Sameera$^{1,5}$, Tom J. L. C. Bakx$^{1}$, Susanne Aalto$^{1}$, Kirsten K. Knudsen$^{1}$, Gunnar Nyman$^{5}$, Wouter Vlemmings$^{1}$ 
\vspace{8pt}}

\begin{document}
\maketitle

\begin{affiliations}
\small
 \item Department of Physics and Astronomy, Chalmers University of Technology, SE-412 96 Gothenburg, Sweden
 \item Univ Rennes, CNRS, IPR (Institut de Physique de Rennes) - UMR 6251, F-35000 Rennes, France
 \item CY Cergy Paris Université, Observatoire de Paris, Université PSL, Sorbonne Université, Université Paris Cité, CNRS, LIRA, F-95000, Cergy, France 
 \item Analytical Mineralogy Group, Institute of Geosciences, Friedrich Schiller University Jena, Jena, Germany
 \item Department of Chemistry and Molecular Biology, University of Gothenburg, Box 462, 40530 Gothenburg, Sweden
\item[\textsuperscript{*}] These authors contributed equally.

\end{affiliations}

\begin{bibunit}
\vspace{-0.5cm}
\begin{abstract}
Cosmic dust is the solid phase of the interstellar medium (ISM), classically assumed to be composed of carbonaceous and silicate grains with size distributions spanning $\mathbf{\sim5~\AA}$ to $\mathbf{\sim1~\mu}$m \cite{WeingartnerDraine2001ApJ...548..296W, DraineLi2007ApJ...657..810D, HensleyDraine2023ApJ...948...55H}. While it constitutes at most order-of-magnitude $\mathbf{1\%}$ of the ISM mass, dust is second only to stars in importance for the observable properties of galaxies \cite{Zavala2021ApJ...909..165Z}. Large uncertainties in the efficiency of grain growth obfuscate the relative contribution of the two dominant sources of dust in the Universe: direct production from evolved stars versus gas-phase accretion in the ambient ISM \cite{Feldmann2015MNRAS.449.3274F, Esmerian2022ApJ...940...74E, Esmerian2024ApJ...968..113E}. Advances in supercomputers have only recently allowed us to move beyond simple, idealized predictions of dust grain growth efficiencies \cite{Leitch-DevlinWilliams1985} with atomistic dynamical calculations \cite{Bossion2024}. We show that small carbon dust grains can grow significantly on timescales much shorter than the age of the universe and, in some ISM phases, comparable to the lifetimes of giant molecular clouds. Specifically, we perform molecular dynamics simulations of an amorphous carbon (a-C) grain surface impacted by gas-phase atoms of cosmologically abundant elements with realistic interstellar conditions, finding high ($\gtrsim 0.2$) sticking coefficients for all non-inert elements at all relevant gas and grain temperatures. We present the results of experiments conducted on similar dust candidate materials that support our theoretical calculations. Our results therefore confirm that the process of gas-phase accretion onto grains is likely an efficient mechanism for the growth of interstellar dust mass on astrophysical timescales, and plausibly central to the evolutionary life-cycle of interstellar grains at all cosmic epochs. Our findings also suggest the presence of non-standard compositions for a-C grains that include elements such as O, Fe, Si, S, Al, and Ni. Consequently, models for the observable properties of interstellar dust may need to allow for a wider range of possible constituent materials than have been considered previously.
\end{abstract}

Carbon-carrying dust is present in the ISM, predominantly of amorphous nature \cite{Mennella1998, Henning2004, Hensley2020ApJ...895...38H, Herrero2022}. Small, i.e. radii $\sim 5$~\AA~to $30$~nm, hydrogenated amorphous carbon (a-C:H) dust is thought to be responsible for the enhanced interstellar extinction around 2175~\AA~\cite{Ysard2024A&A...684A..34Y, HensleyDraine2023ApJ...948...55H}. It is also present as bare amorphous carbon (a-C) in the H-poor environments in which dust grains nucleate from the gas phase \cite{Hecht1991, Crawford2025}, and is likely de-hydrogenated in the intense radiation fields of young, massive stars \cite{Elyajouri2024A&A...685A..76E}. a-C material therefore represents a plausible initial condition and reasonable first approximation for much of the carbonaceous dust in the universe.

Dust grains are thought to nucleate in the relatively warm and very dense environments of stellar remnant ejecta on timescales of days to thousands of years \cite{Sarangi2018SSRv..214...63S, Micelotta2019supe.book..361M, Hofner2018A&ARv..26....1H, Richardson2025ApJ...987..160R}, but they subsequently evolve traversing the interstellar medium for millions to billions of years before being destroyed by high-energy processes, consumed by protostars, or locked in protoplanetary disks, where they provide the seeds of planet formation \cite{Draine2009ASPC..414..453D, Birnstiel2024ARA&A..62..157B}. Estimated grain sizes \cite{WeingartnerDraine2001ApJ...548..296W, HensleyDraine2023ApJ...948...55H} and interstellar gas densities \cite{Draine2011piim.book.....D} suggest that individual grains should collide with much more than their original mass in gas-phase atoms on interstellar, galactic, and cosmological timescales \cite{Draine1990}. Therefore, if impacting gas-phase atoms are incorporated into existing grains with any appreciable probability, this gas-phase accretion is likely a central, possibly dominant, process for determining the dust content of the entire universe \citep[e.g.][]{Dwek1998ApJ...501..643D, Hirashita2000PASJ...52..585H, Feldmann2015MNRAS.449.3274F}. Indirect but highly compelling observational evidence for this process comes from interstellar depletions \cite{Savage1996, Jenkins2009ApJ...700.1299J, RomanDuval2021ApJ...910...95R}, and gas-phase accretion could provide a natural explanation for the high dust masses observed in distant galaxies when the universe was much younger \cite[i.e. $\lesssim 1$ Gyr,][]{Watson2015Natur.519..327W, Michalowski2015A&A...577A..80M, Esmerian2022ApJ...940...74E, Esmerian2024ApJ...968..113E}

\begin{figure*}[ht!]
  \begin{center}
  \includegraphics[width=\linewidth]{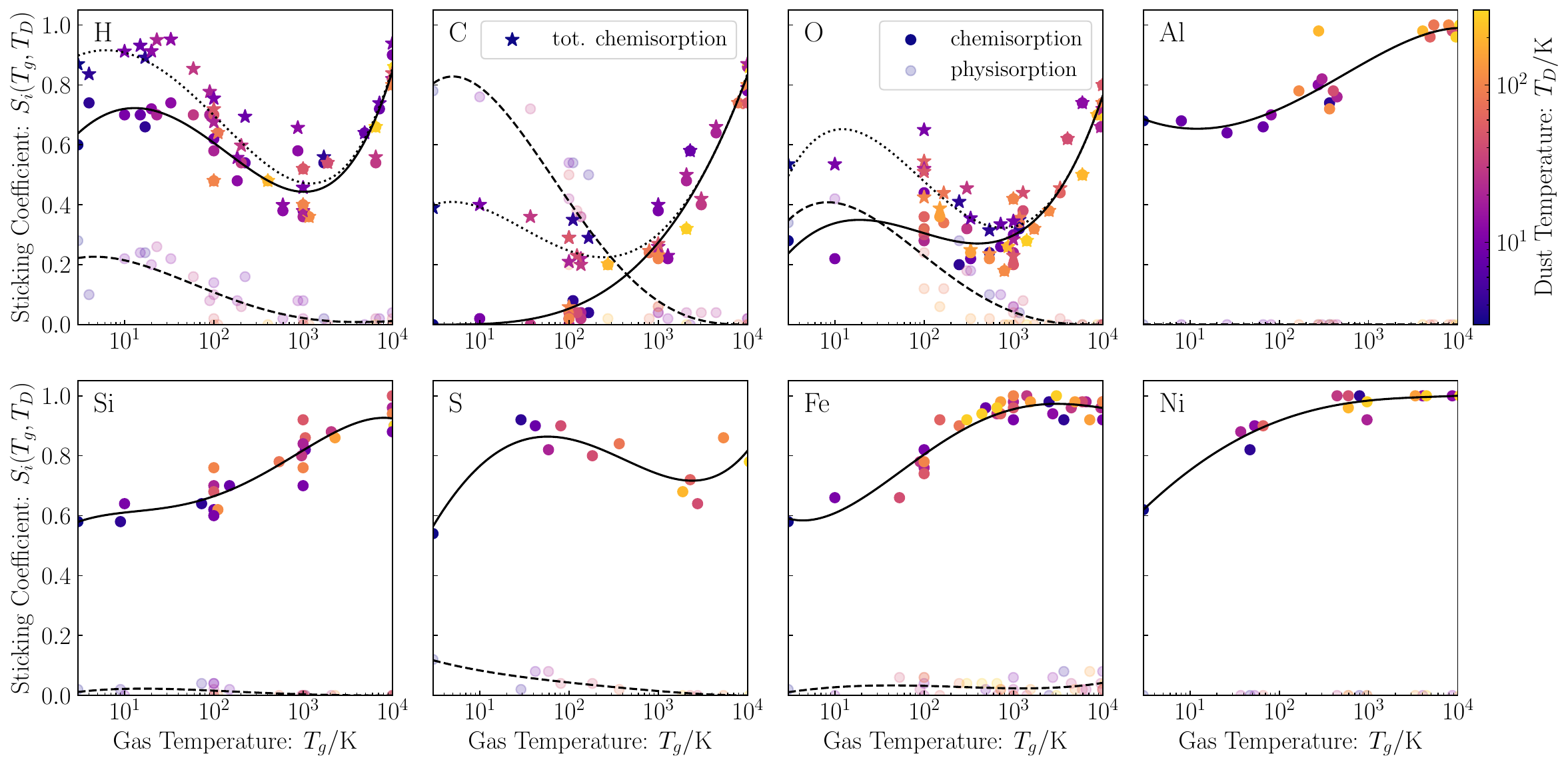}
  \end{center}
    \caption{\small \textbf{Sticking coefficients for selected elements onto amorphous carbon as a function of gas temperature $T_g$ and dust temperature $T_D$. Solid points indicate chemisorption sticking coefficients and transparent points indicate physisorption. Note that this distinction is made at the end of the simulation run-time, and the chemisorption values shown are therefore lower limits (on {\it pure} a-C dust, but see text for discussion on hydrogenation) for astrophysical purposes, since some of the physisorbed atoms may be chemisorbed on longer time-scales. Stars indicate estimated asymptotic total chemisorption values taking the physisorption-to-chemisorption probabilities for H, C, and O from \cite{Bossion2024}. Solid, dashed, and dotted lines indicate polynomial fits -- calculated in $S_i,\;\log_{10}(T_g)$ space -- to the chemisorption, physisorption, and total chemisorption estimates respectively, described further in the Methods.}}
   \label{fig:sticking_coeffs}
\end{figure*}

The rate at which interstellar grains grow due to gas-phase accretion of species $i$ is linearly dependent on the sticking coefficient $S_i(T_g, T_D)$. This is the Maxwell-Boltzmann-velocity-distribution-averaged probability that a particle of species $i$ with gas phase temperature $T_g$ sticks to a dust grain with temperature $T_D$. Sticking coefficients are therefore crucially important to astrophysical models of dust evolution, but have until now been very uncertain: they are not necessarily unity because the impacting atom must establish a covalent chemical bond with -- i.e. chemisorb to -- the grain substrate, the precise dynamics of which will be determined by the complex interactions between the grain surface and the collider. We thereby use molecular dynamics (MD) simulations to estimate sticking coefficients on realistic atomistic models of dust grain surfaces.

Specifically, for the MD interaction potential we use a force field method, which approximates interatomic potentials representing different types of interactions \citep[such as covalent bonds and van der Waals interactions, see e.g.][]{Goddard2008} with functionals that are parametrized to match data obtained by more accurate calculations or experiments. To capture the bond formation and breaking that determines whether the atom will stick to the grain through chemisorption or physisorption (when the atom does not form a covalent bond with the substrate but is held by, e.g., van der Waals bonds), we use the reactive force field (ReaxFF) algorithm \cite{vanDuin2001JPCA}, a method that evaluates the interaction potential between atoms. We propagate the nuclear dynamics classically, as is standard in MD for systems with hundreds of degrees of freedom or more.

\begin{figure*}[ht!]
  \begin{center}
  \includegraphics[width=\linewidth]{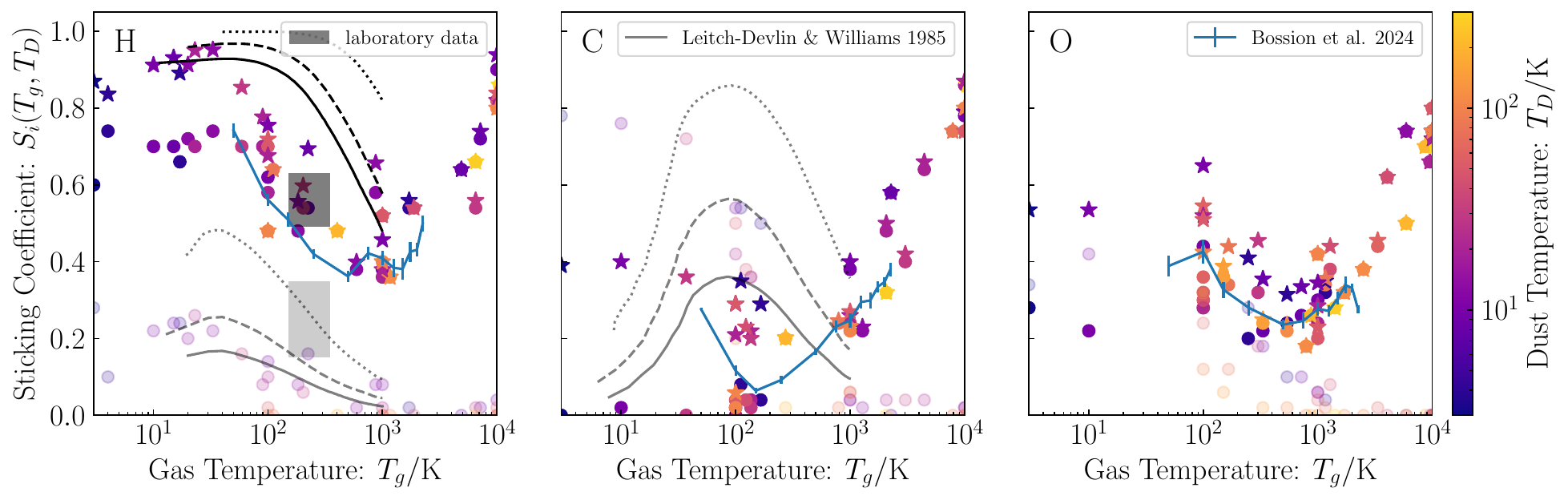}
  \end{center}
    \vspace{-0.6truecm}  
    \caption{\small \textbf{Same as Fig.\ref{fig:sticking_coeffs}, additionally showing comparisons to our laboratory data and previous theoretical estimates. Solid, dashed, and dotted lines indicate previous estimates of sticking coefficients on graphite from \cite{Leitch-DevlinWilliams1985} -- different line-styles indicate grain temperatures of 300, 100, and 3 K respectively, while solid black indicates chemisorption and transparent gray indicates physisorption. For H, the gray rectangles indicate experimental constraints as described in the text: the lighter gray corresponds to dust surface temperatures of both 100 K and 300 K, while the darker gray corresponds to a dust surface temperature of 10 K. Note that these constraints represent lower limits on the {\it combined} chemisorption and physisorption sticking probability. The estimated total chemisorption sticking coefficients from \cite{Bossion2024}, in which $T_g = T_D$,  are shown in blue for comparison.}}
   \label{fig:sticking_coeffs_comp}
\end{figure*}

The MD simulations consist of 50 colliding gas particles on 5 different amorphous carbon structures from various positions and colliding angles around the grain per combination of gas and dust temperature ($T_g,~T_D$). The structures themselves are approximately cubic of size 14~\AA$\times$14~\AA$\times$14~\AA, such that each face represents a section of a larger grain's surface. The dust grain is connected to a thermostat to keep it at $T_D$, and the gas has a kinetic energy related to a velocity-scaled Maxwell-Boltzmann distribution at $T_g$. Further details are presented in the Methods. We perform simulations at selected temperature combinations subject to the following constraints: $T_g \in [3, 10^4]~{\mathrm{K}}$, $T_D \in [3, 300]~{\mathrm{K}}$, and $T_D \le T_g$. The minimum temperature is the lower limit set by the Cosmic Microwave Background \cite[$T_{\mathrm{CMB}} = 2.73$ K,][]{Planck2020}. The maximum gas temperature corresponds to the hottest regions in the dense phases of the ISM \cite{Balser2011ApJ...738...27B}, above which the kinetic energy of incoming atoms becomes comparable to the grain atom binding energies \cite{Bossion2024, Hansson2026A&A...707A..54H} and sputtering may occur. The maximum dust temperature is chosen to encompass the plausible range for dust grains in the ambient interstellar medium, which are almost always observed to be less than 100 K \cite{Paladini2012ApJ...760..149P, BinderPovich2018ApJ...864..136B}.

Chemisorption and physisorption probabilities are calculated from these simulations. Because the binding energies of physisorbed atoms are much lower than those of chemisorbed atoms, we expect only chemisorption is responsible for dust growth on astrophysical time scales. However, some physisorbed atoms will eventually covalently bond to the grain (i.e. chemisorb), and to account for this a physisorption-to-chemisorption probability estimating the asymptotic chemisorption rate for H, C, and O is adopted from our previous work~\cite{Bossion2024}. This is done by extending the simulation time and observing the long-time behavior of physisorbed gas particles. Other elements we study have negligible physisorption rates. 

We calculate the sticking coefficients of several of the most abundant elements in the ISM -- namely H, C, O, Al, Si, S, Fe, and Ni -- onto a pure amorphous carbon surface. Despite Mg, N, and Ca also being relatively abundant in the ISM, we do not calculate their sticking coefficients because force field potentials adapted to the present study are not available for these elements. As a test, we also calculated sticking coefficients for Noble gases He, Ar, and Ne, all of which were equal to zero as expected.

Figure \ref{fig:sticking_coeffs} shows the calculated sticking coefficients as a function of gas-phase and dust grain temperature for the elements considered. For all elements, the total chemisorption sticking coefficient is predicted to be greater than $0.2$ for all gas and dust temperatures. The sticking coefficient appears to be primarily determined by the gas phase temperature for each element, and there is no clear trend with dust temperature. Hydrogen, carbon, and oxygen all display a non-monotonic dependence on gas phase temperature, declining from $T_g \sim 10\;{\mathrm{K}}$ to $T_g\sim 100-1000\;{\mathrm{K}}$ and subsequently increasing to $T_g \sim 10^4\; {\mathrm{K}}$. All other elements display approximately monotonically increasing or constant sticking coefficients with increasing gas temperature, all with values above 0.5. We speculate that the increased sticking coefficients at high temperatures, exhibited by all elements except for S, is due to the increased ability of incoming atoms with higher kinetic energy to dislodge existing chemical bonds and therefore create new free valence electron shells, enabling new covalent bonds to form.

These results are compared to the previous state-of-the-art theoretical calculations and new laboratory estimates in Figure \ref{fig:sticking_coeffs_comp}. The only existing theoretical estimates for similar values were obtained with perturbative quantum mechanical calculations \citep{Leitch-DevlinWilliams1985}, which could only be done assuming a uniform lattice for the grain substrate, i.e. graphite in this case, instead of amorphous carbon. They also could not predict chemisorption of carbon, and did not provide predictions for any elements beyond H and C. Moreover, the authors of \cite{Leitch-DevlinWilliams1985} considered the processes of chemisorption and physisorption independently, leading to total sticking coefficients greater than 1 in some regimes, which is unphysical if physisorption and chemisorption are considered mutually exclusive outcomes. This contradiction can be reconciled if some physisorbed atoms later chemisorb in their modeling, but this not specified in the paper, which makes comparison to their predictions and our data somewhat ambiguous.

Nevertheless, while the results of \cite{Leitch-DevlinWilliams1985} are in decent agreement with ours for H, these predictions qualitatively disagree with the gas temperature dependence of carbon physisorption measured in our simulations, predict a much stronger dust temperature dependence than found in our calculations, and fail to capture our predicted increase in sticking coefficients with gas temperatures above 1000 K. Our calculations are therefore a qualitative and significant improvement over this 4-decades-old, previously state-of-the-art result. Thus, our results should be of general use to the astrophysical dust modeling community. To aid their use, we provide coefficients for polynomials fits for each element in the Methods. We also compare to the results of \cite{Bossion2024}, in which the same calculations as ours were performed but assuming $T_g = T_D$, and we find very good agreement, further confirming the lack of a strong dependence on $T_D$.

To experimentally assess our findings, we obtained recombination coefficients for hydrogen on a carbon dust analog in the laboratory using methods from \cite{Grieco2023NatAs...7..541G, Pirronello1999A&A...344..681P, Hornekaer2003Sci...302.1943H, Aimaud2007msl..confE..71A}, shown as shaded gray regions in Figure \ref{fig:sticking_coeffs_comp}. The recombination coefficient is a lower bound on the sum of the physisorption and chemisorption coefficients, and is therefore not a direct measurement of the sticking coefficient. However, the results are completely consistent with our findings, placing lower limits on $S_{\mathrm{H, chem}} + S_{\mathrm{H, phys}} \ge 0.25 \pm 0.1$ for gas temperatures $T_g = 250 \pm 100\;{\mathrm{K}}$ and dust temperatures $T_D = 100-300\;{\mathrm{K}}$; and $S_{\mathrm{H, chem}} + S_{\mathrm{H, phys}} \ge 0.56 \pm 0.07$ at the same gas temperatures and $T_D = 10\;{\mathrm{K}}$. This consistency with laboratory data gives us some confidence in our numerical results, especially since we expect $S_{\mathrm{H, chem}} + S_{\mathrm{H, phys}} \to S_{\mathrm{H, chem}}$ for the cold temperatures and long timescales of dynamical relevance to the interstellar medium. See Methods for further details.

\begin{figure}[ht!]
  \begin{center}
  \includegraphics[width=\linewidth]{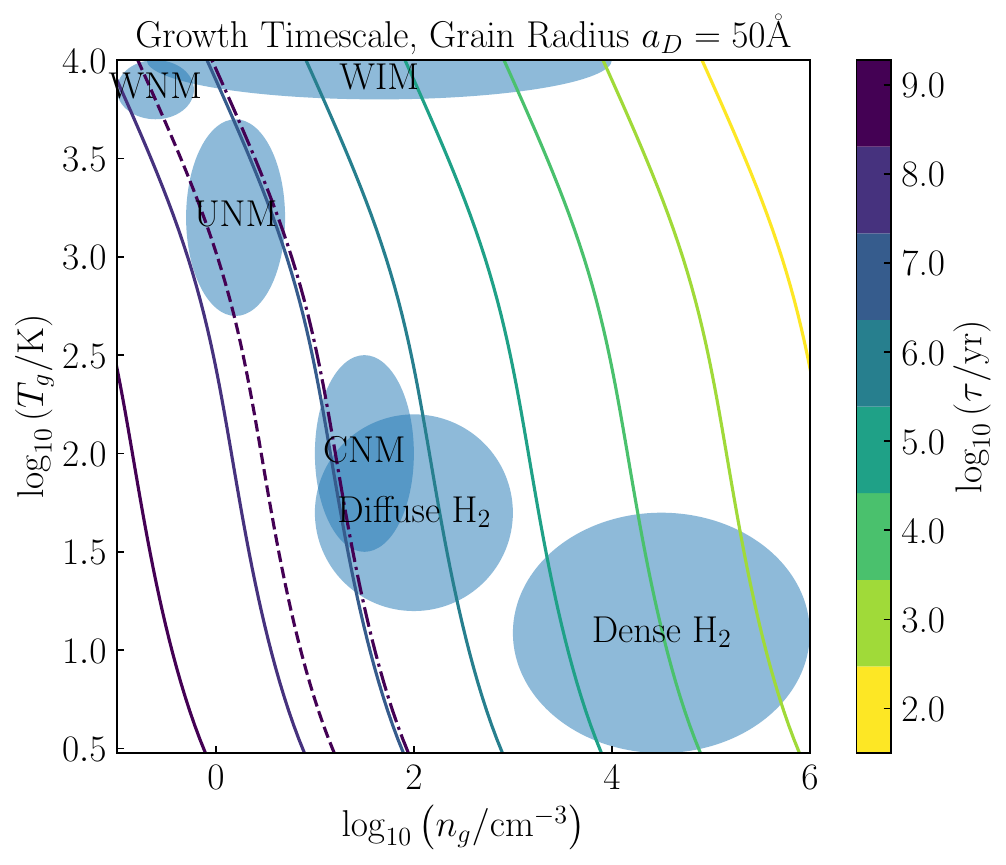}
  \end{center}
    \caption{\small \textbf{Timescales for grain growth via accretion of a $50~\mathrm{\AA}$~radius grain in the different phases of the ISM with our calculated sticking coefficients (Equation \ref{eq:growth_timescale}). Lines show decades in characteristic grain growth timescale from $10^2$ to $10^9$ years as a function of interstellar gas-phase temperature and density. We also show with a dashed line $\tau = 10^9~{\mathrm{yr}}$ for a larger $a_D = 0.1~\mu{\mathrm{m}}$ grain, and with dot-dashed $\tau = 10^9$~yr for timescales multiplied by $10^2$ (offset slightly for visibility) to account for a maximal duty-cycle of de-hydrogenation, see text. Overlaid in blue ellipses are the approximate densities and temperatures of the main thermodynamic phases of dense interstellar gas: the Warm Ionized Medium: WIM, Warm Neutral Medium: WNM, the Unstable Neutral Medium: UNM, the Cold Neutral Medium: CNM, diffuse molecular gas: Diffuse H$_2$, and dense molecular gas: Dense H$_2$ \citep{Draine2011piim.book.....D, Kwok2007pcim.book.....K, HeilesTroland2003ApJ...586.1067H}.}}
  \label{fig:tau_nT}
\end{figure}

These values allow us to estimate the rate of grain-growth due to gas-phase accretion in different interstellar environments. We define the characteristic timescale of grain growth as $\tau \equiv \frac{m_D}{dm_D/dt}$ where $m_D$ is the dust grain mass, for which 

\begin{equation}\label{eq:growth_timescale}
\tau = \frac{\sqrt{2\pi}}{3} \frac{a_{D}\rho_D}{n_g\sqrt{k_{\mathrm{B}}T_g}}\left[\sum_i S_{i}(T_g, T_D) \frac{Z_i}{\mu_i}\sqrt{m_i}\right]^{-1}
\end{equation}

\noindent assuming grains are compact spheres with radius $a_{D}$, material density $\rho_D$ (for which we take 2 g/cm$^3$)\footnote{\url{https://physics.nist.gov/cgi-bin/Star/compos.pl?mode=text&matno=006}}, and temperature $T_D$, accreting from a gas that has number density $n_g$, temperature $T_g$, and is composed of species $i$ with gas mass fractions $Z_i$ \cite[using solar abundances from][]{Draine2011piim.book.....D}, mean molecular weights $\mu_i$, and masses $m_i$ (see Methods). We sum over all elements in Figure ~\ref{fig:sticking_coeffs} except for H, which we discuss below. Sticking coefficients $S_{i}(T_g, T_D)$ are evaluated at arbitrary temperatures using the best-fit polynomials shown in Figure ~\ref{fig:sticking_coeffs} and described in the Methods, and were set to 0 for elements for which we did not measure sticking coefficients (most notably N, Mg, and Ca). This gives timescale estimates as a function of gas phase density and temperature shown in Figure ~\ref{fig:tau_nT} for a grain of radius $a_D=50$~\AA, characteristic of the carbonaceous nano-particles to which the enhanced interstellar opacity at 2175~\AA~is attributed \cite{HensleyDraine2023ApJ...948...55H, Ysard2024A&A...684A..34Y}. Also shown are the $\tau=10^9$~yr curves for a larger grain $a=0.1~\mu$m (dashed), and for an $a_D=50~{\mathrm{\AA}}$ grain increased by a factor of $10^2$ (dot-dashed, discussed below). Overlaid are the characteristic gas densities and temperatures of the thermodynamic phases which compose the dense ISM  \citep{Draine2011piim.book.....D, Kwok2007pcim.book.....K, HeilesTroland2003ApJ...586.1067H}.

Our sticking coefficients predict a wide range of grain-growth timescales in different phases, but are broadly less than 100 Myr for all phases, and as low as $10^4$ to 1000 yrs for the densest phases. Since $\tau \propto a_D$, timescales are proportionally longer for larger grains, shown by the dashed line for $a_D = 0.1~\mu$m, for which timescales are still less than a Gyr for all phases except the most diffuse Warm Neutral Medium (WNM). These timescales are much less than the age of the universe at all but the earliest cosmic epochs \cite[$1-10$ Gyr,][]{Planck2020}, and are comparable to or shorter than best estimates for the other timescales characteristic of gas and grain interstellar processes such as molecular cloud formation and disruption \citep[10--30 Myr,][]{Chevance2020MNRAS.493.2872C, Peters2017MNRAS.467.4322P}, grain nucleation from evolved stellar sources \citep[40 Myr--1 Gyr,][]{SchneiderMaiolino2024A&ARv..32....2S}, and grain destruction due to sputtering in shocks and the hot ionized medium \citep[$\lesssim$ 400 Myr,][]{Draine2009ASPC..414..453D}. This comparison gives substantial support to the possibility that a significant fraction of the cosmic dust mass of the universe may be acquired through gas-phase accretion. These are also consistent with grain growth accretion timescales estimated based on the scaling of dustiness with metallicity in local galaxies \cite{Feldmann2015MNRAS.449.3274F}.

We note that there are three main caveats to this order-of-magnitude estimate of grain growth timescales from gas-phase accretion which should be considered. The first is the possible presence of ice mantles: in the molecular phases of the ISM primarily H$_2$O ice is observed \cite{Gibb2004ApJS..151...35G, Boogert2011ApJ...729...92B, Boogert2015ARA&A..53..541B, Decleir2025AJ....169...99D} and should naturally form on grain surfaces given the high abundance of H and O \cite{Dulieu2010A&A...512A..30D}. Any part of the grain surface covered in ice by more than a few monolayers cannot participate in the growth of the refractory substrate (since the relatively low binding energies of ices will cause them to quickly photodesorb from grain surfaces once they are transported into the diffuse interstellar medium), so ice mantles could present a challenge to growth via gas-phase accretion in the molecular phases of the ISM \cite{Ferrara2016MNRAS.463L.112F}. However, the grain surface covering fraction and mantle thickness of these ices is not known and could be low depending on the structure of the grain surfaces, which may be fractal \cite{Potapov2020PhRvL.124v1103P} or the frequency with which these mantles are disrupted by grain-grain collisions \cite{Tielens1994ApJ...431..321T, Jones1996ApJ...469..740J, Esmerian2026arXiv260522225E}. The effect of ices on grain growth in dense clouds therefore remains an open question necessitating further investigation. However, even if ices are a significant impediment to grain growth via accretion in the molecular phases, there is roughly twice as much mass in more diffuse phases of the local ISM in which ices are not observed \cite{Whittet1997ApJ...490..729W}, which we now consider.

The second major caveat for our prediction of grain growth timescales relates to the Coulomb repulsion between positively charged grains and ions, which is relevant precisely in the diffuse phases of the ISM outside of molecular clouds where the interstellar radiation field is much stronger. While this radiation field prevents the formation and accumulation of ice molecules, it also imparts positive charges to both grains \cite[via the photoelectric effect,][]{WeingartnerDraine2001ApJS..134..263W, Weingartner2006ApJ...645.1188W} and atoms with ionization potentials less than 13.6 eV \cite[i.e. all but O among the elements we consider, via photoionization,][]{Draine2011piim.book.....D}. Equilibrium grain charges from theoretically predicted photoelectron yields do suggest that Coulomb repulsion should dramatically reduce the accretion of ions onto the largest ($a_D \gtrsim 0.01~\mu$m) carbonaceous grains in the Cold Neutral Medium (CNM). However, it is not a prohibitive barrier for smaller grains, or grains of any size in the other phases \cite{WeingartnerDraine1999ApJ...517..292W}. Moreover, the sticking coefficient is likely to be enhanced with ionic species compared to the neutral atoms studied here, possibly having a compensatory effect. All of this is further complicated by the dependence of grain charge on the (very uncertain) photoelectric work function, as well as the radiation field strength and the local interstellar temperature and electron density \cite{WeingartnerDraine2001ApJS..134..263W, IbanezMejia2019MNRAS.485.1220I}. These effects can only be fully understood in the context of a turbulent, multiphase ISM, ideally addressed with 3D fluid-dynamical modeling fully coupled to the evolution of grain population and local radiation field. 

The third caveat regarding our extimated grain growth rates results from the fact that we simulate sticking on pure a-C structures (as opposed to an a-C:H material) which is maximally reactive because hydrogen can fill carbon valence electron shells in a way other carbon atoms cannot. Therefore, our sticking coefficient estimates on a-C are, in principle, upper limits for a-C:H dust. Presently, stable structures for a-C:H at any degree of hydrogenation are not available for use in MD calculations. Our bare a-C grains correspond to the limiting case of low hydrogenation a-C:H, or freshly nucleated pure carbon grains.

 Hydrogenation will therefore likely slow carbon grain growth in the ISM, but not enough to prevent this process from being an important contribution to the total dust mass on astrophysical timescales, because hydrogen has a significantly lower binding energy to amorphous carbon than at least carbon and oxygen \cite{Bossion2024}. Radiative desorption, sputtering from high-energy gas-grain collisions, cosmic ray ionization, high-temperature sublimation, grain-grain collisions, and H$_2$ formation will all therefore preferentially de-hydrogenate grain surfaces, allowing for the accretion of more tightly-bound elements. Indeed, observations of star forming regions suggest the de-hydrogenation of carbon grains by intense radiation fields, leaving bare carbon grains to accrete other elements \cite{Elyajouri2024A&A...685A..76E}. The inefficiency of star formation within individual molecular clouds suggests that much of the interstellar medium is cycled through HII regions many times \cite{Semenov2017ApJ...845..133S}, meaning grains will repeatedly de-hydrogenate and accrete heavier elements over cosmic time. Even if this duty cycle extends the grain-growth timescale by a factor of $10^2$ -- because HII regions last for $\sim$~Myrs \cite{Battersby2017ApJ...835..263B, Bonne2023A&A...679L...5B, Faerber2025ApJ...990...30F} while the orbital timescale for a galactic disk is order-of-magnitude $\sim 100$ Myrs \cite{Cautun2020MNRAS.494.4291C} -- growth timescales for carbon nanograins remain shorter than the Hubble time for all phases, and below 1 Gyr for all but the warm and unstable neutral media (WNM and UNM). Growth is therefore likely still significant over the cosmological timescales on which galaxies accumulate their dust.

We end with a discussion of the potential observational implications of our findings. Our results suggest that ``mixed'' composition carbon-carrying grains with trace amounts of O, Fe, Si, S, Al, and Ni should form readily in the ISM. This would require extending the standard two-species paradigm which modern dust models typically assume: grains are composed primarily of separate amorphous hydrocarbon or amorphous silicate populations, or some composite material with locally silicate or carbonaceous regions in individual grains. This paradigm is motivated in large part by the main spectral features of dust opacities and emissivities: a 2175~\AA~feature, emission lines with wavelengths $\lambda\in(3, 20)~\mu{\mathrm{m}}$ associated with PAHs, and enhanced opacities at 9.7$~\mu{\mathrm{m}}$ and 18$~\mu{\mathrm{m}}$ attributed to amorphous silicates \cite{DraineHensley2021ApJ...909...94D, Ysard2024A&A...684A..34Y}. Moreover, a lack of correlation between the strengths of these features in some datasets has been used to argue that they must come from separate grain populations \cite{Gordon2021ApJ...916...33G, Decleir2025AJ....169...99D}.

However, there are several lines of evidence suggesting a broader diversity of carbon-carrying materials, as our high sticking coefficients predict: laboratory analyses of pre-solar grains in micrometeorites contain silicon carbide \cite[SiC,][]{Stephan2024ApJS..270...27S}, as do observations towards the galactic center \cite{Min2007A&A...462..667M}, and depletion measurements suggest the necessity of iron carbide (FeC) dust especially in low-metallicity environments \cite{RomanDuval2022ApJ...928...90R}. Additionally, the observed gas-phase depletions of O \cite{Whittet2010ApJ...710.1009W}, Fe \cite{Poteet2015ApJ...801..110P}, and S \cite{Yang2024ApJ...974...30Y} in different interstellar environments have not yet been unambiguously explained by astrophysical dust models, and the broader range of possible carbonaceous materials suggested by our results might be explanatory. Generally, the spectral features of dust grains are sensitive to their average bulk composition and trace inclusions of other elements are not expected to have a strong signal in astronomical observations, and therefore cannot easily be ruled out. Theoretical calculations and especially laboratory measurements of the optical properties of composite materials, as well as more constraints on interstellar dust compositions with extinction and depletion measurements, are therefore strongly motivated by our findings. 

\vspace{3pt}
\noindent\rule{\linewidth}{0.4pt}
\vspace{3pt}


{
\color{white}{\cite{Deringer2018,vanDuin2001JPCA,Goddard2008,AMS,RFFSCM,Goddard2008,vanDuin2016,vanDuin2012,Weismiller2010,Kamat2010,Mathews2012,Islam2016,Goddard2010,Grieco2023NatAs...7..541G,Pirronello1999A&A...344..681P, Hornekaer2003Sci...302.1943H, Aimaud2007msl..confE..71A,Potapov2025ApJ...993...49P,Potapov2019ApJ...880...12P,Thrower2012ApJ...752....3T, WANG20122729, harris2020array, Virtanen2020SciPy-NMeth, Hunter:2007}}} 

\putbib[sn-bibliography]

\end{bibunit}

\appendix
\include{methods}
\end{document}

%% file: methods.tex
\begin{bibunit}
\noindent\textbf{\large Methods}

\section{Molecular Dynamics Simulations}

The a-C structures are taken from \cite{Deringer2018}, where they were generated using MD with Gaussian approximation potentials based on machine learning, with a close-to-density-functional-theory accuracy. We selected 5 of their 512 C atom cubic structures for this study, their largest available. The sticking coefficient MD calculations are obtained following two steps, as presented in detail in our previous study on sticking coefficients \cite{Bossion2024}, and using the ReaxFF module of the Amsterdam Modeling Suite \cite{vanDuin2001JPCA,Goddard2008,AMS,RFFSCM}. 

The first step is thermalization to prepare the a-C structures at the desired dust temperature, $T_D$, by connecting them to a Nosé-Hoover chain that acts as a thermal bath and using force field for the dynamics. The force fields used for the calculation of each sticking coefficient are given in Table ~\ref{FF_list}. To ensure that the structure is fully thermalized, the MD is run for 5~ps. 

In the second step, molecular guns are used to shoot ten colliding atoms, one every 1.25~ps, from different positions and at different colliding angles around the a-C structure in order to obtain sticking coefficients that are averaged over many possible surface sites, and not biased by local structures. For the same reason, we repeat this procedure with 5 different a-C structures, so that each sticking coefficient is calculated from 50 total collisions. Each collision is ensured to be independent by continuous thermalization for the duration of each MD simulation.

The initial gas atom velocity is chosen at random from a Gaussian distribution matching as closely as possible a velocity-weighted Maxwell-Boltzmann distribution appropriate for the desired gas temperature, $T_g$. This constraint comes from the inability of the AMS software to assign initial kinetic energies from non-Gaussian distribution functions.

The chemisorption sticking coefficient is obtained by counting the number of binding collisions over the total number of collisions for the chosen gas element, and the physisorption sticking coefficient by counting the number of gas atoms still physisorbed 108.75~ps after the last collision.

\begin{table}
\caption{Force fields used for the molecular dynamics.}
\label{FF_list}
\begin{tabular}{l|r}
    Colliding atom & Reference to the force field used \\
    \hline \hline
    H & CHO\cite{Goddard2008} \\
    C & CHO\cite{Goddard2008} \\
    O & CHO\cite{Goddard2008} \\
    Al & AlCHO\cite{vanDuin2016} \\
    Si & SiC\cite{vanDuin2012} \\
    S & HCONSB\cite{Weismiller2010,Kamat2010,Mathews2012} \\
    Fe & CHFe\cite{Islam2016} \\
    Ni & NiCH\cite{Goddard2010} \\
\end{tabular}
\end{table}

\section{Sticking coefficient from H recombination measurements on porous carbonaceous dust grains}

Experimentally, sticking coefficients are not measured directly. Similarly, it is not possible to distinguish between physisorption and chemisorption for adsorption. Nevertheless, experiments can provide limiting constraints on sticking coefficients. Here, we have studied in particular the recombination of H on carbonaceous grains to constrain the sticking coefficients for this case.

The novel experiments discussed in this article follow the same protocol as  \cite{Grieco2023NatAs...7..541G} for surface temperatures of 10, 100 and 300 K, as well as the more traditional temperature programmed desorption (TPD) methods for 10 K surface temperature case \cite{Pirronello1999A&A...344..681P, Hornekaer2003Sci...302.1943H, Aimaud2007msl..confE..71A}. The notable differences with previous experiments are 1) we used an amorphous, porous carbon substrate, an analogue of carbonaceous cosmic dust, produced in the Jena Dust Machine \cite[e.g.][and references therein]{Potapov2025ApJ...993...49P}. The morphology of the sample can be described as a layer of dust aggregates with sizes of up to several tens of nm having a fractal structure (fractal distribution of grain monomers) and a very high porosity, up to 90\% and 2) the experiments were also performed with H atoms rather than D, unlike previous studies.

The experiments have been performed under ultra-high vacuum (UHV) conditions. The amorphous carbon sample was held at a fixed temperature and exposed to an atomic H beam. The temperature of the beam is estimated to be $T_g = 250 \pm 100$ K.  The beam arrives at an angle of incidence of 45\textdegree~to the normal of the surface, and the number of H$_2$ molecules desorbing from the surface (at an angle of 0\textdegree) is measured by quadrupole mass spectrometry (QMS). The number of incident atoms and molecules formed (once the other components have been subtracted) is calculated, and a recombination efficiency $R$ is calculated, which gives the fraction of atoms sent that emerge from the surface in the form of molecules.

When the surface temperature is significantly higher than 50 K, here we take 100 K and 300 K, we assume that all the molecules formed will desorb in a short time, none being kept on the surface. For these we measure $R= 0.25\pm0.10$. Before forming a molecule, the atoms must necessarily stick to the surface. What we measure is therefore the minimum value of the sticking coefficient (either by physisorption or chemisorption), and $R$ is therefore a lower limit on the sum of sticking by physisorption and chemisorption. We note that it is very likely that sticking by physisorption is underestimated because an atom can stick and desorb quickly without recombining at high temperatures. In the case of a chemisorbed atom, it may not recombine if the newly formed chemical bond with the surface is very stable, gradually hydrogenating the surface, as observed in the case of coronene molecules \cite{Thrower2012ApJ...752....3T}. However this progressive hydrogenation is slow and not very effective, since the addition of H atoms is always in competition with H$_2$ abstraction. The fraction of hydrogenation of the surface is difficult to evaluate quantitatively, so we prefer to adopt the conservative interpretation that $R$ represents a lower limit on the sum of the two sticking pathways.

When the surface temperature is around 10 K, the H$_2$ molecules formed have a significant probability of remaining trapped on the surface. Therefore, in addition to the molecules that desorb during exposure, those that recombine and remain trapped must also be taken into account. To do this, a complementary TPD study must be carried out after a short irradiation.

We found a total $R = 0.56\pm0.07$ at 10 K. The more comprehensive series of experiments, including isotopic effects, will be published in a more detailed article but they indicate that chemisorption is at play, and probably dominant, at all temperatures.

\section{Sticking Coefficient Definition and Relation to Growth Rate}

The growth rate of an individual dust grain with mass $m_D$, temperature $T_D$, and cross section $\sigma_D$ can be expressed as the sum over all gas-phase species $i$ with particle masses $m_i$, gas number densities $n_i$ and impacting velocity distribution function $f(v_i)$ defined by gas temperature $T_g$ as 

\begin{multline}
    \frac{dm_D}{dt} = \sum_i m_i n_i \int_{T_g}dv_i \sigma_{D,i}(m_D, T_D, v_i)v_i f(v_i) \\ = \sum_i m_i n_i\langle\sigma_{D,i}v_i\rangle_{T_g}
\end{multline}

\noindent Without loss of generality, we can decompose the cross section into a geometric $\varsigma_D(m_D)$ component -- which is equal to $\pi a_D^2 \equiv \pi \left(3m_D/4\pi \rho_D\right)^{2/3}$ in the case of a compact sphere with radius $a_D$ and material density $\rho_D$ -- and a dimensionless velocity-and-species-dependent component $s_i(v_i, T_D)$ which represents the probability of a particle with velocity $v_i$ sticking to the grain. This gives

\begin{equation}
    \frac{dm_D}{dt} = \sum_i m_i n_i \varsigma_D  \langle v_i\rangle S_i(T_g, T_D)
\end{equation}

\noindent where

\begin{equation}
    \langle v_i\rangle S_i(T_g, T_D) \equiv \int_{T_g}dv_i v_i f(v_i)s_i(v_i, T_D)
\end{equation}

\noindent defines the sticking coefficient $S_i(T_g, T_D)$ and 

\begin{equation}
    \langle v_i\rangle = \sqrt{\frac{8 k_B T_g}{\pi m_i}}
\end{equation}

\noindent assuming a Maxwell-Boltzmann velocity distribution function of impacting atoms from the gas phase. The grain accretion rate assuming a compact spherical geometry is then given by 

\begin{equation}
    \frac{dm_D}{dt} = a_D^2 \sum_i S_i(T_g, T_D) n_i  \sqrt{8\pi k_B T_g m_i} .
\end{equation}

By expressing the species number density as a function of the total gas number density $n_g$

\begin{equation}
    n_i = \frac{Z_i}{\mu_i}n_g
\end{equation}

\noindent where $Z_i$ is the mass-fraction abundance for element $i$ and $\mu_i \equiv m_i/m_{\rm H}$, and the dust grain mass as $m_D = 4\pi a_D^3 \rho_D/3$, we obtain the characteristic growth timescale in Equation \ref{eq:growth_timescale}.

\section{Sticking Coefficient Fitting Functions}\label{methods:fits}

We fit functions of the form

\begin{equation}\label{eq:fitting_func}
f(x) = \sum_{k=0}^4 \beta_k b_k(x, 4)
\end{equation}

\noindent where $b_k$ is the k-th order Bernstein polynomial

\begin{equation}
b_k(x, N) = \frac{N!}{k!(N - k)!}x^k(1 - x)^{N - k}.
\end{equation}

\noindent and $x$ represents the logarithm of the gas phase temperature $T_g$ scaled as                                                                        

\begin{equation}
x \equiv \frac{\log_{10}(T_g/T_{g,{\rm min}})}{\log_{10}(T_{g,{\rm max}}/T_{g,{\rm min}})}                                                                      
\end{equation}

\noindent so that the domain of the data is $x\in[0, 1]$, which ensures $f\in[0,1]$ over the same domain if $\beta_k$ are also constrained to the interval $[0,1]$, which is ensured by a constrained least-squares fit of Eq.~\ref{eq:fitting_func} to the data \cite{WANG20122729}. This procedure is used to obtain the fitting functions in Figure~\ref{fig:sticking_coeffs}, which are then used to estimate chemisorption sticking coefficients of arbitrary gas-phase temperature needed for the calculation of Figure~\ref{fig:tau_nT}. The coefficients $\beta_k$ for each fitting function are provided in Table~\ref{fit_coeffs}.

\begin{table*}
\caption{Constants for sticking coefficient fitting functions of form given in Equation~\ref{eq:fitting_func} for $T_g \in [3, 10^4]$ K.}
\label{fit_coeffs}      
\begin{tabular}{|c|c|c|c|c|c|}
\hline
Element, Sticking Type & $\beta_0$ & $\beta_1$ & $\beta_2$ & $\beta_3$ & $\beta_4$\\
\hline
H chem. & 0.637692 & 0.880471 & 0.642446 & 4.26551$\times10^{-9}$ & 0.844555\\
H phys. & 0.222050 & 0.266711 & 6.89822$\times10^{-11}$ & 8.04638$\times10^{-10}$ & 1.16518$\times10^{-2}$\\                                                    
H est. tot. chem. & 0.897642 & 1.000000 & 0.717216 & 9.71200$\times10^{-12}$ & 0.852628\\                                                                       
C chem. & 1.36215$\times10^{-9}$ & 1.47347$\times10^{-13}$ & 4.96691$\times10^{-14}$ & 0.131301 & 0.832154\\                                                    
C phys. & 0.804555 & 1.000000 & 3.17827$\times10^{-2}$ & 0 & 9.53020$\times10^{-12}$\\                                                                          
C est. tot. chem. & 0.399147 & 0.490532 & 8.60753$\times10^{-11}$ & 0.112538 & 0.837005\\                                                                       
O chem. & 0.236891 & 0.516770 & 0.250330 & 1.42055$\times10^{-12}$ & 0.762474\\
O phys. & 0.347339 & 0.611140 & 0 & 0 & 3.20174$\times10^{-14}$\\
O est. tot. chem. & 0.495774 & 1.000000 & 0.231138 & 2.05038$\times10^{-12}$ & 0.761758\\                                                                       
Al chem. & 0.686203 & 0.589352 & 0.687057 & 0.976380 & 0.989953\\
Al phys. & 0 & 0 & 0 & 0 & 0\\
Si chem. & 0.579043 & 0.670899 & 0.508266 & 0.961495 & 0.924595\\
Si phys. & 1.14554$\times10^{-2}$ & 4.23356$\times10^{-2}$ & 3.61731$\times10^{-3}$ & 0 & 4.01880$\times10^{-12}$\\                                             
S chem. & 0.564342 & 1.000000 & 1.000000 & 0.496617 & 0.816421\\
S phys. & 0.117419 & 4.77560$\times10^{-2}$ & 4.36416$\times10^{-2}$ & 3.45520$\times10^{-14}$ & 4.60292$\times10^{-12}$\\                                      
Fe chem. & 0.590406 & 0.519715 & 1.000000 & 1.000000 & 0.959414\\
Fe phys. & 1.07340$\times10^{-2}$ & 4.89992$\times10^{-2}$ & 3.84052$\times10^{-2}$ & 3.60950$\times10^{-12}$ & 4.19600$\times10^{-2}$\\                        
Ni chem. & 0.619440 & 0.868705 & 1.000000 & 0.988224 & 1.000000\\
Ni phys. & 0 & 0 & 0 & 0 & 0\\
\hline

\end{tabular}
\end{table*}

\vspace{3pt}
\noindent\rule{\linewidth}{0.4pt}
\vspace{3pt}

\let\oldthebibliography=\thebibliography
\let\oldendthebibliography=\endthebibliography
\renewenvironment{thebibliography}[1]{%
    \oldthebibliography{#1}%
    \setcounter{enumiv}{73} 
}{\oldendthebibliography}
%

%
%
\begin{addendum}
 \item[Acknowledgements]  C.J.E., D.B., W.M.C.S., T.J.L.C.B., S.A., K.K., G.N., and W.V. acknowledge financial support from the Knut and Alice Wallenberg foundation through grant no. KAW 2020.0081. The computations were enabled by resources provided by the National Academic Infrastructure for Supercomputing in Sweden (NAISS), partially funded by the Swedish Research Council through grant agreement no. 2022-06725. This work was funded by CY Initiative of Ex-cellence (grant ``Investissements d'Avenir'' ANR-16-IDEX-0008). AP acknowledges support from the Deutsche Forschungsgemeinschaft (Heisenberg grant PO 1542/7-1 and research grant PO 1542/12-1). This article made use of the following software packages: NumPy \cite{harris2020array}, SciPy \cite{Virtanen2020SciPy-NMeth}, Matplotlib \cite{Hunter:2007}. This research has made use of NASA's Astrophysics Data System Bibliographic Services.

\item[Author Information] The authors declare that they have no competing financial interests. Correspondence and requests for materials should be addressed to C.J.E.~(email: clarke.esmerian@chalmers.se) and D.B.~(email: duncan.bossion@univ-rennes.fr).

\item[Author Contributions] C.J.E. conceptualized the project, helped plan the MD simulations, performed the analysis necessary for and made the figures, led the scientific interpretation, and led writing the initial text. D.B. planned and performed the MD simulations, performed their analysis for the calculation of sticking coefficients, contributed to the scientific interpretation, and contributed to writing the initial text including but not limited to all sections describing MD simulation methods and analysis. F.D., S.B., and A.P. conceived of, set up, and performed the laboratory experiments, and F.D. contributed the initial text describing their methods and results. W.M.C.S. facilitated the laboratory experimental collaboration, helped plan the MD simulations, and contributed to the scientific interpretation. T.J.L.C.B. contributed to the scientific interpretation and chose the title. S.A., K.K., G.N., W.V. all contributed to the scientific interpretation and led securing computational resources and funding for the project. All authors contributed to manuscript revision and finalization.

\item[Data availability] The datasets generated during and/or analyzed in the present study are available from the corresponding authors on reasonable request. 
 
\item[Code availability]  The analysis scripts used for the current study are available from the corresponding authors on reasonable request.
\end{addendum}



\end{bibunit}